\documentclass[letterpaper]{article} 
\usepackage{aaai24}  
\usepackage{times}  
\usepackage{helvet}  
\usepackage{courier}  
\usepackage[hyphens]{url}  
\usepackage{graphicx} 
\usepackage{natbib}  
\usepackage{caption} 
\usepackage{fancyhdr}

\usepackage{algorithm}
\usepackage{algorithmic}

\usepackage{newfloat}
\usepackage{listings}
\DeclareCaptionStyle{ruled}{labelfont=normalfont,labelsep=colon,strut=off} 
\floatstyle{ruled}
\newfloat{listing}{tb}{lst}{}
\floatname{listing}{Listing}
\usepackage{subfiles}
\usepackage{xcolor}
\title{Position Paper on Materials Design -- A Modern Approach}
\author{
    Willi Gro{\ss}mann\textsuperscript{\rm 1}, 
    Sebastian Eilermann\textsuperscript{\rm 1}, 
    Tim Rensmeyer\textsuperscript{\rm 1}, \\
    Artur Liebert\textsuperscript{\rm 1},  Michael Hohmann\textsuperscript{\rm 1}, Christian Wittke\textsuperscript{\rm 1},
    Oliver Niggemann\textsuperscript{\rm 1}
}
\affiliations{
\textsuperscript{\rm 1}Institute of Automation Technology\\Helmut-Schmidt-Universität / Universität der Bundeswehr Hamburg, \\
Holstenhofweg 85, 22043 Hamburg, Germany\\
}

\usepackage{bibentry}

\begin{document}

\maketitle
\thispagestyle{fancy}

\begin{abstract}
Traditional design cycles for new materials and assemblies have two fundamental drawbacks. The underlying physical relationships are often too complex to be precisely calculated and described. Aside from that, many unknown uncertainties, such as exact manufacturing parameters or materials composition, dominate the real assembly behavior. Machine learning (ML) methods overcome these fundamental limitations through data-driven learning. In addition, modern approaches can specifically increase system knowledge. Representation Learning allows the physical, and if necessary, even symbolic interpretation of the learned solution. In this way, the most complex physical relationships can be considered and quickly described. Furthermore, generative ML approaches can synthesize possible morphologies of the materials based on defined conditions to visualize the effects of uncertainties. This modern approach accelerates the design process for new materials and enables the prediction and interpretation of realistic materials behavior.

\end{abstract}

\section{Introduction}
\label{Introduction}

The traditional design of new materials and their production are based on exploratory research and heuristic methodology \cite{morgan2020opportunities,butler2018machine,MolMatSurv}. Newer approaches use computer-aided modeling of materials properties and integrate them into the materials design process. This enables a more efficient screening process for possible theoretical solutions \cite{ramprasad2017machine, sousa2021material}. But materials properties are still described spatially using physical models. These models are always based on an idealized theoretical morphology. 

This is where state-of-the-art laboratory-based computed tomography (CT) can create CTs of the real morphology. It is an accurate and fast way to digitize materials and assemblies morphologies over statistically relevant volumes and with appropriate resolution \cite{tom3, tom2, tom1}. Once a digital image of the morphology is available, superimposed numerical solutions of the partial differential equations can approximate the effective materials properties \cite{num2, num1}. However, numerical solving is associated with an enormous computational effort. In addition, uncertainties in the theoretical description with physical models, for example due to the system complexity or the non-ideal behavior of some components, remain undescribed.

Data-driven methods close the gap between empirical observation and physical modelling. Neural networks have become the dominant approach for data-based modeling in many complex areas \cite{YOLO,UNET, AlphaGo}. Their success is largely due to their ability as universal function approximators and the existence of highly efficient optimization algorithms \cite{Optimizers}. However, in many neural network architectures, the purely data-based approach limits their ability to extrapolate to data that is very different from the training data \cite{IPNN}. Nevertheless, neural networks are successfully used to create models for material properties \cite{MLMF2,MLMF1, MatMLSurv,MolMatSurv}.

Unfortunately, the black-box nature of neural networks means that the learned function is usually uninterpretable by a human. This makes it difficult to use the trained model to gain deeper insights into the physical principles that lead to the observed properties of the materials, or even to assess the reliability of the model prediction. As a response to the aforementioned shortcomings of standard neural networks, several alternative approaches have been developed that improve these models by learning physical concepts of varying degrees of interpretability. \cite{PIM, GAN, Graph}

One promising method in this domain is representation learning (RepL). The aim of this field is to create models which are capable of mapping observations of a system into representations that are useful for making downstream predictions for the system \cite{RepL}. These representations usually offer an improved degree of interpretability and are often explicitly designed for that purpose \cite{kingma:2014a, BVAE, Eureqa, AIFeynman}. A number of research hypotheses (RH) arise from this context. The proposed solution takes these into account and enables a faster and more efficient material design process.

\textit{RH 1:} The solution is based on a formalization of materials designs that capture the relevant materials properties and serve as input to a surrogate model. Such a model must be developed based on requirements and uncertainties from manufacturing processes.

\textit{RH 2:} The network typologies (Bayesian Generative Autoencoders) and the corresponding algorithms (active learning, RepL) must be developed for this domain. Neural networks for time series -- which would be suitable -- are an on-going research field in ML \cite{HEWAMALAGE2021388,8489399}. Here too, a training strategy must be developed.

\textit{RH 3:} To extract an explicit, formula-based material model from the neural network, several research directions exist: From symbol-regression based-approaches \cite{doi:10.1073/pnas.1517384113}, via questions-based algorithms \cite{PhysRevLett.124.010508} to approaches which define priors on latent variables \cite{kingma:2014a}. Here a suitable algorithm and neural network type must be identified and verified.

\textit{RH 4:} Materials morphologies need to be encoded into a set of feature vectors that are automatically extracted from CT data. Some of these features are straight-forward (e.g., density), while others will require novel image-processing strategies. In related applications, convolutional neural networks have proven to be particularly suitable for this.

\textit{RH 5:} CT data can be used to quantify the effects of manufacturing processes. This knowledge enables high-resolution conditional generative neural networks (generative adversarial networks (GANs) \cite{goodfellow2014generative} or denoising diffusion probabilistic models (DDPM) \cite{ho2020denoising}) to synthesize CTs based on material and manufacturing parameters. For this purpose, a suitable experimental design must be created. In addition, the algorithms must consider highly modal and diverse conditions.

\section{State of the Art}
\label{SoA}
\begin{figure*}[ht]
    \centering
    \includegraphics[width=0.95\textwidth]{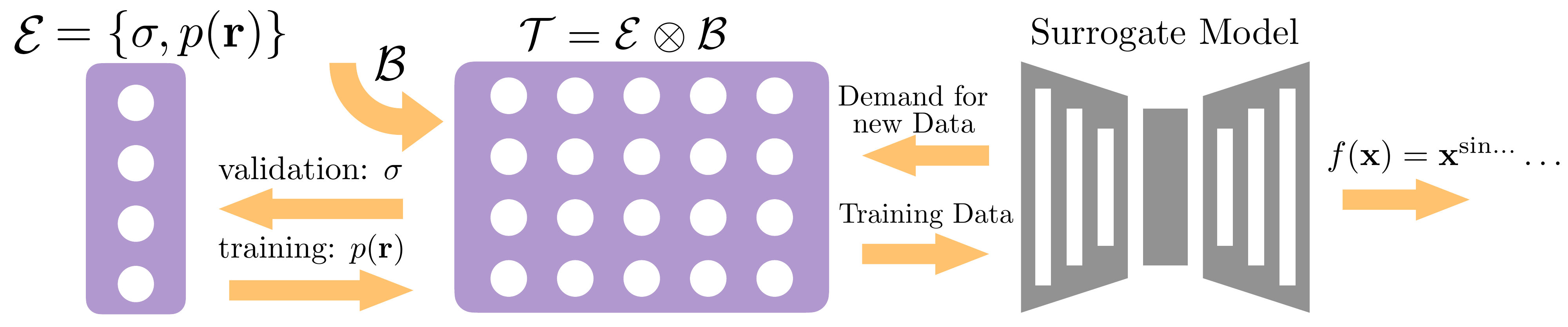}
    {\caption{\small Tiered strategy to learn materials laws; a targeted experimental program will generate realistic morphologies $\bf r$ and reference data $\sigma$. Direct physical modeling merges experimental data ${\cal E}$ with bulk data ${\cal B}$ from databases to generate large scale feature-property data ${\cal T}$. This is then used to learn a surrogate model in form of a Bayesian generative neural network autoencoder. This autoencoder can both in an active learning sense request new data points. But it can also be a basis for the generation of a symbol model, e.g. using algorithms from the field of explainable AI.}\label{fig:workflow}}%
\end{figure*}
The author group traditionally works on the edge between computer science and automation technology. Focus are ML and artificial intelligence methods for cyber-physical systems including RepL \cite{Niggemao3, Niggemao4}. 
Considering a-priori knowledge to improve ML is one big research topic. Therefore, technical systems are characterized by the abundance of a-priori knowledge such as system structures and physical laws. As a result, methods of machine learning require less data and become extrapolable. Furthermore, causal relationships are used to calculate optimizations \cite{Niggemao6, Niggemao1, Niggemao2}. These partly subsymbolic representations are used to abstract logical symbols from the physical systems \cite{Niggemao8}. 
to generate surrogate models \cite{Niggemao7}.

For several years now we have been abstracting these skills into the sub-area of computational materials design or -- from an algorithmic perspective -- spatial neural networks. Often neural networks are used to analyze materials, e.g. to identify characteristics \cite{Jung-Niggemann,grossmann:2022h}.

This also includes the use of surrogate models to replace slow solver-based simulations when designing new materials \cite{rensmeyer:2023a}
 with generative and probabilistic neural networks \cite{Niggemann:2023b}.

\subsection{Physics Based Machine Learning}
The ideal physical behavior of materials can often be approximated using well-known partial differential equations (PDEs). However, the numerical solution of these equations is computationally intensive for high temporal and spatial resolutions. In addition, classic deep learning approaches fail due to a lack of labeled data. Physics-informed neural networks (PINNs) solve this contradiction \cite{Cai.2021, BPINN, PINNFlow}.

These models learn the unknown parameters in an otherwise known (partial) differential equation. To do this, the neural network is trained to interpolate between the known data points in a way that corresponds to the known form of the equation. The unknown parameters of the equation are treated as neural network parameters and optimized by gradient descent alongside the actual neural network parameters \cite{RAISSI}.


An alternative to PINNs for settings where the governing equations are not already partially known is offered by an uncertainty-based active learning strategy. They have established themselves as an extremely valuable approach in areas where labeling data requires significant effort \cite{DAL}. These methods have proven capable of substantially reducing the amount of required data to train a neural network model, by only labeling data on which the model still has a high degree of uncertainty \cite{active}. Additionally, this approach makes the resulting model more robust to outliers, as exploiting uncertainty in the labeling process can be used to bias the dataset toward outlier configurations \cite{UAviaAL}. 

\subsection{Generative Systems Modelling}
An alternative for the spatial system description, taking complex (boundary) conditions into account, are generative AI approaches such as GANs or DDPMs. CTs help here as training data to create realistic morphologies. 

For example, GANs are used to generate a large number of CT images that have similar properties to the training images \cite{su2022microstructure}. 
In other approaches, the corresponding three-dimensional (3D) structure is derived from two-dimensional (2D) images \cite{FENG2020113043,ZHANG2021110018}.
\citet{Henkes_2022} proposes an extended method that uses a Wasserstein loss \cite{arjovsky2017wasserstein} to generate 3D microstructures with the same properties as the original data. 

However, to create morphologies with defined and changing properties, conditions-based approaches must be considered. Although there are already some approaches from the field of medical technology \cite{Jiang2020COVID19CI,Peng2023CBCTBasedSC,CganMed1} and materials design \cite{DURETH2023105608,janssens2020computed}, the underlying conditions are not multimodal or high-dimensional.

Feature engineering methods help with subsequent automated CT segmentation or classification, as well as design evaluation \cite{damewood2023representations}. Convolutional Neural Networks (CNNs) are used successfully here \cite{GOBERT2020101460,ziabari2023enabling}. However, a disadvantage of this approach is the poor interpretability of the learned features.

The aim of these intermediate steps is a closed ML framework that covers the entire materials design process \cite{10.3389/fmats.2022.818535,pei2021machine}. The list of the previous approaches shows the possibility for abstract and idealized descriptions of materials behavior. However, since only the idealized materials behavior is described there, uncertainties regarding production and composition are not taken into account. Furthermore, explicit, interpretable, or formula-based substitution models are missing.

\subsection{Systems Representation}
RepL is the process of automatically discovering and extracting meaningful patterns and features from raw data, enabling more effective and efficient machine learning models. Autoencoders and variational autoencoders \cite{kingma:2014a} have become highly utilized methods for the extraction of physically meaningful representations of a systems observations. Nautrup et al. and Iten et al. \cite{PhysRevLett.124.010508, Nautrup} use these methods to identify the relevant variables of a physical system that are necessary for downstream computations and representation in a simple and interpretable way. Variational autoencoders have also been used to improve the interpretability of latent variables by producing discrete encodings of observations \cite{MixReps, SOMVAE, VQVAE}. The learning of discrete representations has been one of the main focuses of RepL and could potentially be used to identify different modes of behavior of materials properties. 

Champion et al. \cite{AutoSR} use an autoencoder to map the observation of the physical system to a suitable representation.  A symbolic regression of the extracted features is then performed. The symbolic form describes the desired properties of a physical system and have varying degrees of complexity depending on the chosen basis. In doing so, they describe the transition into the field of system identification. The algorithmic formulas are derived from observed data to model the behavior of dynamic systems.

\section{Solution}
\label{Solution}

\begin{figure*}[ht]
\centering
\includegraphics[width=\textwidth]{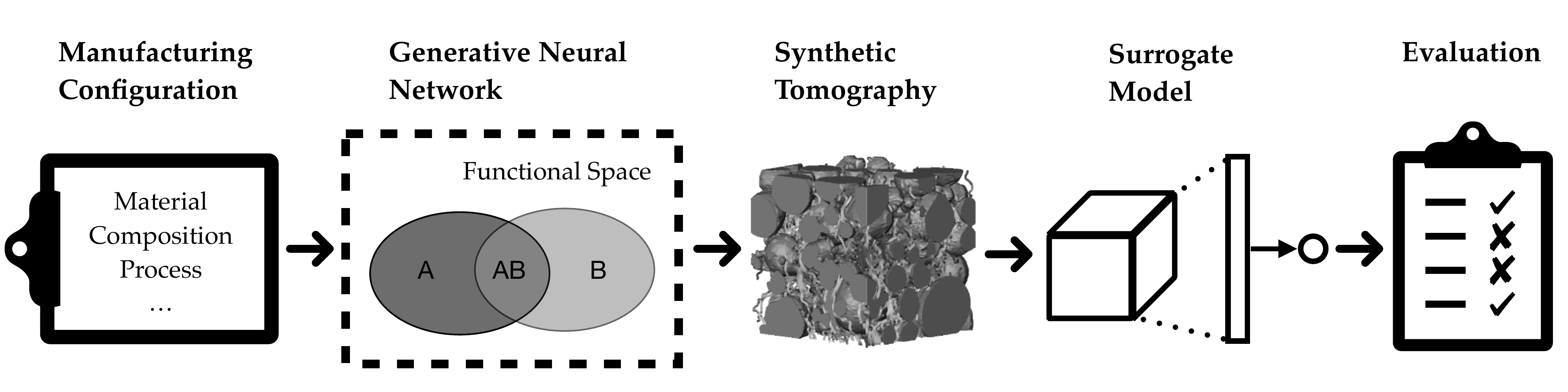}
\caption{Closed framework for morphological system description. A generative neural network synthesizes possible tomographies (CTs) of material composites or assemblies based on predefined manufacturing configurations. The morphological representation enables the visualization of various manufacturing influences. A surrogate model is based on the spatial representation and describes the physical properties. This finally enables an evaluation of the materials design}
\label{fig:Framework}
\end{figure*}
ML, CT, and experiments are combined here, to automatically construct transferable and interpretable constitutive materials laws, which are currently missing for the design of many complex material composites. As a result, a new method is developed to learn surrogate models for materials and assemblies. 

\subsection{Surrogate Design Model}
\label{Surrogate_Model}
A bottleneck in traditional materials design is the time consuming design-test-cycle: Each new material has to be designed, produced and then experimentally tested. Direct simulation is only a partial solution, as both modeling and simulation tasks are time-consuming, require a high level of expertise and significant computational resources to solve even small domains. A solution for the problem is a machine learning surrogate model which is a simplified and computationally less expensive model that serves as a substitute for time-consuming simulations or expensive computations.

Based on bulk materials data, a model generation function can be learned that maps new materials designs onto materials models. Such predictions can be used to try other, new materials designs. Maybe better: Even if this neural network-based approach yields less quantitative precision for a single material. Unbiased learning of a predictive model can still be expected to be qualitatively more predictive than traditional ad hoc heuristics.

Such an approach is sketched in Figure~\ref{fig:workflow}: 
The strategy is three-fold. A targeted experimental program generates a range of materials from which CTs are synthesized to create a data set of typical morphologies. The synthesised materials are then tested for their properties to generate a validation dataset ${\cal E}$, while the CTs are feed into the second layer of physical modelling. The main purpose of this layer is to efficiently expand the training set by overlaying the morphologies with available bulk properties to simulate the behavior of a much larger set of materials. This wouldn't be possible experimentally. The layer thoroughly maps the relationship between bulk properties and morphologies. These are expressed by feature vectors build from materials properties (e.g., porosity $\epsilon_p$, tortuosity $\tau$, strut lengths $l$ etc.)  It also implements a map $p({\bf r})\rightarrow \left\{\epsilon_p,\tau,...\right\}$ through image-processing techniques, which encodes CTs into feature vectors. This large-scale dataset underpins the training phase. Based on this data -- augmented with and validated against real experiments -- a neural network is trained. Later during the operation phase, new materials designs -- expressed by the feature vectors and materials properties -- become input to this network which then predicts the probable corresponding materials model. 

\subsection{Algorithmic Representation}
\label{Algortith_Repr}
Because these materials models come in form of a neural network, explainability and traceability become challenges. For this reason, explicit, formula-based models are extracted from the neural network. Such models resemble closely current manual materials models but can qualitatively differ as a function of provided feature vectors. However, to prevent neural networks from remaining black boxes, changes to the network topology and learning process are required. For this (see also Figure \ref{fig:workflow}) a surrogate model in form of a neural network is learned. This model replaces the previous physical simulations but has the advantage of predicting the materials behavior very quickly. Using such a surrogate model has several advantages: (i) Its topology can be optimized towards the symbolic formula extraction, e.g. by disentanglement of latent variables (RepL). (ii) By using Bayesian principles of uncertainty estimation, the surrogate model can actively ask for more data points when needed. (iii) The surrogate model can be created in a way which supports inter- and extrapolation -- mainly by adding physical information to the network topology and the priors on latent variables. Finally, symbolic formula extraction algorithms and algorithms from the field of explainable AI are used to extract symbolic formulas. These learned formulas guide the design process, e.g. by helping to identify optimal materials designs for specific tasks or identifying promising candidates for targeted experiments.

\subsection{Morphological Representation}
\label{Morph_Repr}
Due to a symbolic description, the physical materials behavior can be theoretically interpreted. However, the generated equations are often difficult to interpret in practice due to their complexity. Another way to better understand system relationships is to synthesize materials structures in the form of tomography. This approach is visualized in Figure \ref{fig:Framework}.

The generative part uses CTs as training data. These are labeled with manufacturing configurations such as material compositions and process parameters. Under these high-dimensional conditions, a GAN or DDPM is trained to synthesize high-resolution 3D tomography data. In the operation phase, the respective influence on the morphological structure is analyzed by varying the conditions. Analogous to the unsupervised approach mentioned above, a surrogate model is then used to learn and represent the materials behavior. Finally, the design draft is evaluated.

\section{Conclusion}
\label{Conclusion}

A comprehensive approach that integrates ML, CTs and experiments is presented. This solves the lack of transferable and interpretable materials laws for complex composite materials and assemblies. The surrogate design model provides an efficient way to overcome the traditional bottleneck in the design-test cycle. 

The algorithmic representation ensures the interpretability of the neural network-based models. The extraction of explicit, formula-based models enhances understanding and facilitates meaningful contributions to materials design. 

In addition, the morphological representation provides a visual understanding of the materials system through synthesis of 3D tomography data. This enables the analysis of manufacturing influences on morphological structures and provides valuable insights into the material design process.

The proposed framework not only streamlines the materials design process but also empowers researchers with a powerful tool to explore and predict the behavior of complex material composites. By combining CTs synthesis and explainable AI, this modern approach opens opportunities for the rapid and informed design of materials with tailored properties.

\section{Acknowledgment}
\label{Ackn}
This work was supported by the project "KIBIDZ --Intelligente Brandgefahrenanalyse f\"ur Geb\"aude und Schutz der Rettungskrafte durch K\"unstliche Intelligenz und Digitale Brandgeb\"audezwillinge" by the Zentrum f\"ur Digitalisierungs- und Technologieforschung der Bundeswehr (dtec.bw®).

\bibliography{lit}

\end{document}